\documentclass[fleqn,usenatbib]{mnras}

\usepackage{newtxtext,newtxmath}

\usepackage[T1]{fontenc}
\usepackage{ae,aecompl}

\usepackage{graphicx}	
\usepackage{amsmath}	
\usepackage{amssymb}	
\usepackage{subcaption}
\usepackage{pdflscape}
\usepackage{multirow}
\usepackage{rotating}

\usepackage{color}
\usepackage{soul}
\usepackage{lipsum}

\title[SCUBA-2 overdensities associated with candidate protoclusters]{SCUBA-2 overdensities associated with candidate protoclusters selected from \textit{Planck} data}

\author[T. Cheng et al.]{
T. Cheng,$^{1}$\thanks{E-mail: t.cheng15@imperial.ac.uk}
D.L. Clements,$^{1}$
J. Greenslade,$^{1}$
J. Cairns$^{1}$
P. Andreani,$^{2}$
M. Bremer,$^{3}$
\newauthor{
L. Conversi,$^{4}$
A. Cooray,$^{5}$
H. Dannerbauer,$^{6,7}$
G. De Zotti,$^{8}$
S. Eales,$^{9}$
}
\newauthor{
J. Gonz\'{a}lez-Nuevo,$^{10, 11}$
E. Ibar,$^{12}$
L. Leeuw,$^{13}$
J. Ma,$^{5}$
M. J. Micha\l owski,$^{14}$
}
\newauthor{
H. Nayyeri,$^{5}$
D. A. Riechers,$^{15, 16}$
D. Scott,$^{17}$
P. Temi,$^{18}$
M. Vaccari,$^{19, 20}$
}
\newauthor{
I. Valtchanov,$^{21}$
E. van Kampen$^{2}$
and
L. Wang$^{22, 23}$
}
\\
$^{1}$Astrophysics Group, Imperial College London, Blackett Laboratory, Prince Consort Road, London SW7 2AZ, UK\\
$^{2}$European Southern Observatory, Karl-Schwarzschild-Stra${\ss}$e 2, 85748 Garching, Germany\\
$^{3}$HH Wills Physics Laboratory, University of Bristol, Tyndall Avenue, Bristol, BS8 1TL, UK\\
$^{4}$European Space Agency / ESAC, Camino Bajo del Castillo, 28692 Villanueva de la Ca\~nada (Madrid), Spain\\
$^{5}$Department of Physics and Astronomy, University of California, Irvine, CA 92697, USA\\
$^{6}$Instituto de Astrof\'{i}sica de Canarias, E-38205 La Laguna, Tenerife, Spain\\
$^{7}$Universidad de La Laguna Dpto. Astrof\'{i}sica, E-38206 La Laguna, Tenerife, Spain\\
$^{8}$INAF-Osservatorio astronomico di Padova, Vicolo dell'Osservatorio 5, I-35122 Padova, Italy\\
$^{9}$School of Physics and Astronomy, Cardiff University, The Parade, Cardiff CF24 3AA, UK\\
$^{10}$Departamento de F\'{i}sica, Universidad de Oviedo, C. Federico Garc\'{i}a Lorca 18, 33007 Oviedo, Spain\\
$^{11}$Instituto Universitario de Ciencias y Tecnologias Espaciales de Asturias (ICTEA), C. Independencia 13, 33004 Oviedo, Spain\\
$^{12}$Instituto de F\'isica y Astronom\'ia, Universidad de Valpara\'iso, Avda. Gran Breta\~na 1111, Valpara\'iso, Chile\\
$^{13}$College of Graduate Studies, University of South Africa, PO Box 392, UNISA, 0003, South Africa\\
$^{14}$Astronomical Observatory Institute, Faculty of Physics, Adam Mickiewicz University, ul.~S{\l}oneczna 36, 60-286 Pozna{\'n}\\
$^{15}$Department of Astronomy, Cornell University, Space Sciences Building, Ithaca, NY 14853, USA\\
$^{16}$Max-Planck-Institut f\"ur Astronomie, K\"onigstuhl 17, D-69117 Heidelberg, Germany\\
$^{17}$Department of Physics and Astronomy, University of British Columbia,Vancouver, BC V6T1Z1, Canada\\
$^{18}$Astrophysics Branch, NASA Ames Research Center, Moffett Field, CA 94035, USA\\
$^{19}$Department of Physics and Astronomy, University of the Western Cape, Private Bag X17, Bellville 7535, Cape Town, South Africa\\
$^{20}$INAF - Istituto di Radioastronomia, via Gobetti 101, 40129 Bologna, Italy\\
$^{21}$Telespazio Vega UK for ESA, European Space Astronomy Centre, Operations Department, E-28691 Villanueva de la Ca\~nada, Spain\\
$^{22}$SRON Netherlands Institute for Space Research, Landleven 12, 9747 AD, Groningen, The Netherlands\\
$^{23}$Kapteyn Astronomical Institute, University of Groningen, Postbus 800, 9700 AV, Groningen, The Netherlands
}

\date{Accepted 2020 April 14. Received 2020 March 18; in original form 2020 January 23}

\pubyear{2020}

\begin{document}
\label{firstpage}
\pagerange{\pageref{firstpage}--\pageref{lastpage}}
\maketitle

\begin{abstract}

We measure the 850-$\mu$m source densities of 46 candidate protoclusters selected from the \textit{Planck} High-z catalogue (PHz) and the \textit{Planck} Catalogue of Compact Sources (PCCS) that were followed up with \textit{Herschel}-SPIRE and SCUBA-2. This paper aims to search for overdensities of 850-$\mu$m sources in order to select the fields that are most likely to be genuine protoclusters. Of the 46 candidate protoclusters, 25 have significant overdensities ($>$5 times the field counts), 11 have intermediate overdensities (3--5 times the field counts) and 10 have no overdensity ($<$3 times the field counts) of 850-$\mu$m sources. We find that the enhanced number densities are unlikely to be the result of sample variance. Compared with the number counts of another sample selected from \textit{Planck}'s compact source catalogues, this [PHz+PCCS]-selected sample has a higher fraction of candidate protoclusters with significant overdensities, though both samples show overdensities of 850-$\mu$m sources above intermediate level. Based on the estimated star-formation rate densities (SFRDs), we suggest that both samples can efficiently select protoclusters with starbursting galaxies near the redshift at which the global field SFRD peaks ($2 < z < 3$). Based on the confirmation of overdensities found here, future follow-up observations on other PHz targets may greatly increase the number of genuine DSFG-rich clusters/protoclusters.

\end{abstract}

\begin{keywords}
galaxies: high-redshift -- galaxies: starburst -- submillimetre: galaxies
\end{keywords}



\section{Introduction}

Protoclusters of galaxies are structures that are expected to collapse into galaxy clusters by $z=0$, but that have yet to fully collapse at the observed epoch \citep{2016A&ARv..24...14O}. They are not yet virialised and so cannot be efficiently found using traditional galaxy cluster detection methods through X-ray emission or the Sunyaev-Zeldovich effect \citep[SZE,][]{1980ARA&A..18..537S}, which require the presence of hot ($10^{7}$--$10^{8}$ K) gas, or through red sequence galaxies \citep{2000AJ....120.2148G}. Current optical/near-infrared (NIR) surveys aiming to detect protoclusters, such as the Hyper Suprime-Cam Subaru Strategic Program \citep[HSC-SSP,][]{2010MNRAS.409.1155D, 2018PASJ...70S..12T}, mainly study overdensities of Lyman-break galaxies (LBGs), Lyman-$\alpha$ emitters (LAEs) or H-$\alpha$ emitters (HAEs) with blind (unbiased) searches or around ``biased tracers" such as QSOs or radio galaxies \citep{2000A&A...361L..25P, 2004A&A...428..817K, 2008A&A...491...89V, 2011PASJ...63S.415T, 2012ApJ...757...15H, 2013MNRAS.432.2869H, 2015ApJ...808L..33C}. Such surveys have found hundreds of candidate protoclusters, but they are unlikely to recover the full protocluster population. Furthermore, optical/NIR surveys miss protoclusters whose member galaxies are havily dust-obscured, which is especially the case at $z>2$.

If we look at the cores of $z \sim 0$ galaxy clusters, there is an abundance of elliptical galaxies \citep{1980ApJ...236..351D, 1988ARA&A..26..509B, 2003MNRAS.346..601G}. According to some galaxy formation models, these elliptical galaxies are the successors of dusty star-forming galaxies (DSFGs) at high redshifts \citep{2006ApJ...641L..17F, 2006ApJ...650...42L, 2008ApJ...689L.101F, 2010MNRAS.402.2113C, 2011ApJ...742...24L, 2013MNRAS.431..648W, 2013ApJ...768...21C, 2014ApJ...782...69L, 2014ApJ...782...68T, 2015ApJ...810...74A, 2017MNRAS.464.1380W}. Adding the fact that protoclusters are the progenitors of $z=0$ galaxy clusters suggests that there should also be an abundance of DSFGs in protoclusters at high redshifts, which is supported by observations \citep{2009ApJ...691..560C, 2009ApJ...694.1517D, 2014A&A...570A..55D, 2015ApJ...815L...8U, 2015ApJ...808L..33C, 2015ApJ...812...43B, 2019ApJ...872..117G}.

Even though there are already observations of protoclusters containing DSFGs, the sample is small compared to that of optical/NIR protocluster surveys. \cite{2005MNRAS.358..869N} developed a technique to detect protoclusters based on their FIR/submm emission. They proposed to use the fact that the FIR flux density in a low-resolution survey is the sum of many sources if they are clustered with a size similar to the beam.  Following this technique, a number of studies have aimed at selecting protoclusters containing DSFGs using the \textit{Planck} Early Release Compact Source Catalog \citep[ERCSC,][]{2011A&A...536A...7P}, the Catalogue of Compact Sources \citep[PCCS,][]{2014A&A...571A..28P}, and the Second Planck Catalogue of Compact Sources \citep[PCCS2,][]{2016A&A...594A..26P}. These studies have produced a sample of candidate protoclusters \citep{2013A&A...549A..31H, 2014MNRAS.439.1193C, 2018MNRAS.476.3336G} and follow-up observations have been obtained \citep[][]{2016MNRAS.461.1719C, 2019arXiv190908977C}. \cite{2018MNRAS.476.3336G}, in particular, estimated the surface density of DSFG-rich candidate protoclusters to be $(3.3 \pm 0.7) \times 10^{-2} deg^{-2}$, consistent with other studies \citep{2014MNRAS.439.1193C}.


\cite{2017MNRAS.468.4006M} (hereafter M17) selected a number of candidate protoclusters with DSFGs from the \textit{Planck} high-z source candidates list \citep[PHz,][]{2016A&A...596A.100P}, the PCCS, and follow-up \textit{Herschel}-SPIRE observations \citep{2015A&A...582A..30P}. M17 also completed follow-up observations of 46 candidate protoclusters using SCUBA-2 at 850 $\mu$m, obtaining photometric redshifts, FIR luminosities, and star-formation rate density (SFRD) distributions. They found that their 850-$\mu$m sample has a redshift peak between $z=2$ and 4, a typical FIR luminosity of $10^{13} \textrm{L}_{\odot}$, an SFRD peak at $z \simeq 3$, and with an uncorrected number density of all sources in the candidate protoclusters being 6 times more than in the field.

Among these 46 candidate protoclusters, two (PLCK$\_$G006.1+61.8 and PHz$\_$G173.9+57.0, see Table \ref{Table_todd}) were also identified as the most overdense candidates in the \textit{Spitzer Planck Herschel} Infrared Cluster survey \citep[SPHerIC,][]{2018A&A...620A.198M} sample, which selects candidate clusters at $1.3 < z < 3$ using photometric data from \textit{Planck}, \textit{Herschel} and \textit{Spitzer}/IRAC.

This paper extends the work of M17. Using their sample of 46 candidate protoclusters, we calculate each of their 850-$\mu$m source densities, classify them based on the derived source densities, and look for the candidate protoclusters that are most overdense in 850 $\mu$m sources. We also compare the source densities with those of 850 $\mu$m observations of candidate protoclusters discussed in \cite{2019arXiv190908977C} (hereafter C19), which were originally selected from the ERCSC, PCCS or PCCS2, and compare the two samples in terms of efficiency of selecting genuine protoclusters.

In Section \ref{todd_selection} we present the selection of candidate protoclusters and our source extraction using SCUBA-2 data. In Section \ref{todd_ncounts} the 850-$\mu$m source densities are shown. We discuss our results and conclude in Section \ref{Discussion}. Unless otherwise stated, we use the standard concordance cosmology with $H_{0} = 67.4 \textrm{km} \textrm{s}^{-1} \textrm{Mpc}^{-1}$, $\Omega_{\textrm{M}} = 0.3$ and $\Omega_{\Lambda} = 0.7$ \citep{2018arXiv180706209P}.

\section{Candidate protocluster selection, source extraction}\label{todd_selection}

Forty-six candidate protoclusters were selected and studied in M17. They were originally selected from the PHz and the PCCS catalogues, with colour cuts using their 857, 545, 353 and 217 GHz flux densities in order to remove cold Galactic cirrus and extragalactic radio sources. According to M17, only sources with an infrared excess, or $S_{545}/S_{857} > 0.5$ and $S_{353}/S_{545} < 0.9$ in the PHz catalogue, and $S_{857}/S_{545} < 1.5$ and $S_{217}/S_{353} < 1$ in the PCCS catalogue, were selected, where $S$ is the flux density.

Among these [PHz+PCCS]-selected sources, 228 were followed up with \textit{Herschel}-SPIRE. Fifteen of these 228 sources were then idenfied as being gravitionally-enhanced submillimetre sources (GEMS), the so-called \textit{Planck} dusty GEMS \citep{2012ApJ...753..134F, 2012A&A...538L...4C, 2015A&A...581A.105C}. After excluding the Galactic cirrus sources, the rest show overdensities of \textit{Herschel}-SPIRE sources with flux densities peaking at 350 or 500$\mu$m \citep{2015A&A...582A..30P}. These sources are believed to be either high-\textit{z} protoclusters or chance line-of-sight projections.

Sixty-one sources were observed with SCUBA-2. Ten of these are GEMS in \cite{2015A&A...581A.105C}, with peak flux densities of 350 to 1140 mJy at 850 $\mu$m. Forty-six are believed to be protoclusters due to their \textit{Herschel}-SPIRE overdensities.

We extract the SCUBA-2 850$\mu$m sources in the same way as in C19. We start from the highest S/N pixel in the S/N map and go down to the detection threshold of S/N=3.5. Connected pixels that have S/N$\geq$3.5 are regarded as part of the same source. As discussed in C19, the detection threshold of 3.5$\sigma$ is chosen since the reliabilility is found to be above 80$\%$ at 3.5$\sigma$ (c.f. C19). The flux density and noise are recorded at the position of the pixel with the highest S/N within a source. Note that the flux density and noise are also deboosted following C19 and  \cite{2017MNRAS.465.1789G} (hereafter Ge17), and a 5$\%$ calibration uncertainty is also included. This equivalently gives our source catalogue a minimum noise value of 1.04 mJy.

The 850-$\mu$m source catalogue is essentially the same as that of M17, though in M17 they applied further constraints to exclude sources that are below signal-to-noise ratios of 4, and sources having 850-$\mu$m flux density uncertainties above 4 mJy. In order to compare with the number count results in C19, we retain our 850-$\mu$m source catalogue\footnote{The source catalogue can be downloaded from the online supplementary material.}, which follows the source extraction method in C19.

We test the completeness of our 850-$\mu$m sources for each candidate protocluster by inserting artificial sources from 2 to 20 mJy into the flux density maps and use the same extraction method. The shapes of these artificial sources are approximated by 2D Gaussians with standard deviations as the SCUBA-2 beamsize at 850 $\mu$m. In order to minimise the chances of sources overlapping, 10 sources are inserted in the map each time, and the process is repeated 1,500 times. We could not totally rule out overlappings of these artificial sources, given their extended 2D Gaussian shapes. Nonetheless, 10 sources is representative of the number of real sources in each map, so our artificial sources should have similar statistical characteristics to the real sources. Given the limiting map sizes, there are also chances of pixel repetitions when inserting these artifical sources for 1,500 times. Nonetheless, the repetitive pixels do not change the noise characteristics and thus the completeness level of each field, so do not change the conclusions made in this paper.

The fraction of extracted and inserted artifical sources in each candidate protocluster, as a function of flux density, is the completeness. Out of the 46 candidate protoclusters, 25 have completeness above 50$\%$, and 21 are below 50 $\%$, at 8 mJy. We mark these $< 50 \%$ completeness candidates in the last column in Table \ref{Table_todd} as ``C". Out of these $< 50 \%$ completeness candidates, 18 still have significant or intermediate overdensities (category II or III as discussed in Section \ref{todd_ncounts}), indicating they are likely to be genuine protoclusters with overdensities of 850 $\mu$m sources even under lower completeness. Out of the $< 50 \%$ completeness candidates, three have no overdensity of 850-$\mu$m sources; they might still have overdensities of 850-$\mu$m sources, but their lower completeness makes it difficult to confirm this.

We found that the higher rms in the flux density maps may explain the low completeness in some fields. The rms values for those $<$50$\%$ completeness fields at 8 mJy are on average twice the rms of the rest. At brighter flux densities, such as at 12 mJy, the number of candidate protoclusters having completeness level $<$50$\%$ decreases to two.

The reliability of our 850-$\mu$m sources for each candidate protocluster is tested by inverting the flux density maps, following M17. After the maps are inverted, ``negative" sources are extracted using the same method as for positive sources. Assuming that the negative sources are due to noise spikes and hence there should be the same number of ``positive" noise spikes, the fraction of these negative sources and positive sources for each candidate protocluster, as a function of S/N, is therefore a measure of reliability. Out of the 46 candidate protoclusters, 32 have reliability above 80$\%$ at 3.5$\sigma$, and 14 have reliability below 80$\%$ at 3.5$\sigma$. We mark these $< 80 \%$ reliability candidates in the last column of Table \ref{Table_todd} as ``R". Among those $< 80 \%$ candidates, there are 11 that show significant or intermediate overdensities of 850-$\mu$m sources (category II or III as discussed in Section \ref{todd_ncounts}). We note with caution that such overdensities might not be real, due to their lower reliability.

\section{Number Counts of [PHZ+PCCS]-selected Candidate Protoclusters}\label{todd_ncounts}



We follow the cumulative number count analysis in C19 using the SCUBA-2 850-$\mu$m source catalogue of the 46 candidate protocluster fields studied in M17. The number of sources is counted cumulatively from brighter to fainter flux density bins with binwidths of 2 mJy. Since the sensitivity varies across the map, we correct the number counts by dividing the number of sources by the effective area corresponding to different sensitivities (rather than the entire map area).

The cumulative number counts of the 46 candidate protoclusters from M17 are shown in Table \ref{Table_todd}. We quote the cumulative number counts from 4 to 12 mJy, which includes the majority of the sources, scaled to the area of each map (of approximately 0.03 deg$^{2}$) and with the variable sensitivity corrected over the map. We estimate the probability $P_{\textrm{ran}}$ of detecting the observed number of sources in each candidate protocluster at 6 or 8mJy, assuming that the sources are randomly distributed and following a Poisson distribution\footnote{$P_{\textrm{ran}}$ is the \textit{upper tail} of the probability density function, following a Poisson distribution. The \texttt{R} function \texttt{ppois(observed-1, lambda=expected, lower=FALSE)} is used to calculate $P_{\textrm{ran}}$.}, by comparing to the field results in Ge17.

\begin{table*}
\centering
\scriptsize
\caption{Cumulative number counts of 46 candidate protoclusters in M17, scaled to the size of each map and with variable sensitivity corrected, following C19. $P_{\textrm{ran}}$ is the probability of detecting the observed number of SCUBA-2 sources compared with the expected number in Ge17 (as shown in the last line) from S2CLS, assuming Poisson statistics. $N_{\textrm{over}}$ is the number of overdense regions (more than observed at 8mJy) when examining 10,000 random regions in the S2CLS/COSMOS field. $P_{\textrm{over}}$ is the probability of obtaining $N_{\textrm{over}}$ regions, i.e. $N_{\textrm{over}}$/10,000. In the last column the sources are classified into three categories, where category I means having no overdensity of SCUBA-2 sources, category II is corresponds to having an intermediate overdensity, and category III is having a significant overdensity, as discussed in Section \protect\ref{todd_ncounts}. For fields below 80$\%$ reliability at 3.5$\sigma$, we add an ``R" label. For fields below 50$\%$ completeness, we add a ``C" label (see Section \protect\ref{todd_selection} for details). The probabilities for PLCK$\_$HZ$\_$G173.9+57.0 are calculated based on the cumulative number counts at 6 mJy, since it does not have any sources brighter than 8 mJy.}
\label{Table_todd}
\begin{tabular}{lccccccccc}
\hline \multicolumn{1}{c}{Name}                                   & $>$4 mJy      & $>$6 mJy              & $>$8 mJy               & $>$10 mJy     & $>$12 mJy     & \begin{tabular}[c]{@{}c@{}}$P_{\textrm{ran}}$ \\ (at 8 mJy)\end{tabular} & $N_{\textrm{over}}$ & $P_{\textrm{over}}$ & Category \\ \hline
Planck18p194                                               & 28$\pm$0.6    & 28$\pm$0.6            & 7.1$\pm$0.07           & 4.3$\pm$0.04  & 3.0$\pm$0.03  & 4.18$\times 10^{-3}$                                                     & 515                 & 0.0515              & II       \\
Planck18p735                                               & 24$\pm$0.6    & 24$\pm$0.6            & 1.1$\pm$0.01           & 1.1$\pm$0.01  & N/A           & 0.86                                                                     & 4540                & 0.454               & I (R)    \\
Planck24p194                                               & 17$\pm$0.5    & 17$\pm$0.5            & 6.3$\pm$0.08           & N/A           & N/A           & 0.02                                                                     & 700                 & 0.07                & II       \\
PLCK$\_$DU$\_$G045.7--41.2                                 & 8.4$\pm$0.1   & 8.4$\pm$0.1           & 8.4$\pm$0.1            & 4.5$\pm$0.04  & 1.1$\pm$0.01  & 9.97$\times 10^{-4}$                                                     & 365                 & 0.0365              & II (R)   \\
PLCK$\_$DU$\_$G059.1--67.1                                 & 9.8$\pm$0.1   & 9.8$\pm$0.1           & 9.8$\pm$0.1            & 6.2$\pm$0.05  & 3.0$\pm$0.03  & 2.13$\times 10^{-4}$                                                     & 242                 & 0.0242              & III      \\
PLCK$\_$DU$\_$G073.4--57.5                                 & 29$\pm$1.5    & 29$\pm$1.5            & 2.7$\pm$0.02           & 2.7$\pm$0.02  & 1.1$\pm$0.01  & 0.59                                                                     & 2883                & 0.2883              & I (R)    \\
PLCK$\_$G006.1+61.8                                        & 12$\pm$0.4    & 12$\pm$0.4            & 12$\pm$0.4             & 12$\pm$0.4    & 10$\pm$0.1    & 1.17$\times 10^{-6}$                                                     & 87                  & 0.0087              & III (C)  \\
PLCK$\_$G009.8+72.6                                        & 42$\pm$2.3    & 42$\pm$2.3            & 13$\pm$0.2             & 7.0$\pm$0.07  & 2.0$\pm$0.02  & 1.75$\times 10^{-7}$                                                     & 31                  & 0.0031              & III (R)  \\
PLCK$\_$G056.7+62.6                                        & 13$\pm$1.2    & 13$\pm$1.2            & 13$\pm$1.2             & 3.7$\pm$0.04  & 3.7$\pm$0.04  & 1.75$\times 10^{-7}$                                                     & 31                  & 0.0031              & III (C)  \\
PLCK$\_$G068.3+31.9                                        & 21$\pm$5      & 21$\pm$5              & 21$\pm$5               & 9.6$\pm$0.6   & 7.1$\pm$0.2   & 4.58$\times 10^{-15}$                                                    & 0                   & $< 10^{-4}$         & III (RC) \\
PLCK$\_$G075.1+33.2                                        & 9.0$\pm$0.6   & 9.0$\pm$0.6           & 9.0$\pm$0.6            & 9.0$\pm$0.6   & 4.8$\pm$0.08  & 2.13$\times 10^{-4}$                                                     & 365                 & 0.0365              & III (RC) \\
PLCK$\_$G077.7+32.6                                        & 9.6$\pm$1.4   & 9.6$\pm$1.4           & 9.6$\pm$1.4            & 6.2$\pm$0.1   & 2.7$\pm$0.03  & 2.13$\times 10^{-4}$                                                     & 242                 & 0.0242              & III (C)  \\
PLCK$\_$G078.9+48.2                                        & 4.7$\pm$0.08  & 4.7$\pm$0.08          & 4.7$\pm$0.08           & 4.7$\pm$0.08  & 2.6$\pm$0.03  & 0.14                                                                     & 1331                & 0.1331              & I (C)    \\
PLCK$\_$G082.5+38.4                                        & 23$\pm$7.0    & 23$\pm$7.0            & 9.1$\pm$0.2            & 4.5$\pm$0.04  & 3.1$\pm$0.03  & 2.13$\times 10^{-4}$                                                     & 242                 & 0.0242              & III      \\
PLCK$\_$G083.3+51.0                                        & 31$\pm$59     & 31$\pm$59             & 12$\pm$0.1             & 12$\pm$0.1    & 4.5$\pm$0.04  & 1.16$\times 10^{-7}$                                                     & 87                  & 0.0087              & III (C)  \\
PLCK$\_$G091.9+43.0                                        & 17$\pm$63     & 17$\pm$63             & 17$\pm$63              & 9.1$\pm$0.2   & 9.1$\pm$0.2   & 4.46$\times 10^{-11}$                                                    & 0                   & $< 10^{-4}$         & III (C)  \\
PLCK$\_$G093.6+55.9                                        & 5.6$\pm$0.07  & 5.6$\pm$0.07          & 5.6$\pm$0.07           & 5.6$\pm$0.07  & 3.2$\pm$0.03  & 0.05                                                                     & 956                 & 0.0956              & I (C)    \\
PLCK$\_$G132.9--76.0                                       & 2.2$\pm$0.02  & 2.2$\pm$0.02          & 2.2$\pm$0.02           & 2.2$\pm$0.02  & 2.2$\pm$0.02  & 0.59                                                                     & 2883                & 0.2883              & I (C)    \\
PLCK$\_$G144.1+81.0                                        & 6.9$\pm$0.3   & 6.9$\pm$0.3           & 6.9$\pm$0.3            & 6.9$\pm$0.3   & 1.1$\pm$0.01  & 0.02                                                                     & 700                 & 0.07                & II (RC)  \\
PLCK$\_$G160.7+41.0                                        & 29$\pm$17     & 29$\pm$17             & 29$\pm$17              & 20$\pm$0.7    & 3.9$\pm$0.06  & 5.85$\times 10^{-24}$                                                    & 0                   & $< 10^{-4}$         & III (C)  \\
PLCK$\_$G162.1--59.3                                       & 37$\pm$1.5    & 37$\pm$1.5            & 3.5$\pm$0.04           & 1.0$\pm$0.01  & 1.0$\pm$0.01  & 0.32                                                                     & 1937                & 0.1937              & I        \\
PLCK$\_$G165.8+45.3                                        & 20$\pm$3.5    & 20$\pm$3.5            & 20$\pm$3.5             & 6.6$\pm$0.1   & 4.8$\pm$0.06  & 4.90$\times 10^{-14}$                                                    & 0                   & $< 10^{-4}$         & III (C)  \\
PLCK$\_$G173.8+59.3                                        & 8.6$\pm$0.7   & 8.6$\pm$0.7           & 4.6$\pm$0.07           & 3.3$\pm$0.03  & 2.2$\pm$0.02  & 0.14                                                                     & 1331                & 0.1331              & I        \\
PLCK$\_$G177.0+35.9                                        & 21$\pm$0.9    & 21$\pm$0.9            & 7.4$\pm$0.08           & 4.4$\pm$0.04  & 2.0$\pm$0.02  & 4.18$\times 10^{-3}$                                                     & 515                 & 0.0515              & II (R)   \\
PLCK$\_$G179.3+50.7                                        & 23$\pm$0.3    & 23$\pm$0.3            & 9.4$\pm$0.08           & 2.1$\pm$0.02  & 2.1$\pm$0.02  & 2.13$\times 10^{-4}$                                                     & 242                 & 0.0242              & III      \\
PLCK$\_$G186.3--72.7                                       & 15$\pm$0.9    & 15$\pm$0.9            & 8.8$\pm$0.1            & 4.8$\pm$0.05  & N/A           & 9.97$\times 10^{-4}$                                                     & 365                 & 0.0365              & II       \\
PLCK$\_$G186.6+66.7                                        & 18$\pm$3.1    & 18$\pm$3.1            & 18$\pm$3.1             & 8.6$\pm$0.1   & 4.2$\pm$0.04  & 4.85$\times 10^{-12}$                                                    & 0                   & $< 10^{-4}$         & III (C)  \\
PLCK$\_$G188.6--68.9                                       & 30$\pm$0.8    & 30$\pm$0.8            & 21$\pm$0.2             & 4.2$\pm$0.04  & 2.0$\pm$0.02  & 4.58$\times 10^{-15}$                                                    & 0                   & $< 10^{-4}$         & III      \\
PLCK$\_$G191.3+62.0                                        & 9.9$\pm$0.4   & 9.9$\pm$0.4           & 9.9$\pm$0.4            & 9.9$\pm$0.4   & 7.4$\pm$0.1   & 2.13$\times 10^{-4}$                                                     & 242                 & 0.0242              & III (C)  \\
PLCK$\_$G191.8--83.4                                       & 34$\pm$0.8    & 34$\pm$0.8            & 20$\pm$0.2             & 3.3$\pm$0.03  & 1.0$\pm$0.01  & 4.90$\times 10^{-14}$                                                    & 0                   & $< 10^{-4}$         & III      \\
PLCK$\_$G201.1+50.7                                        & 23$\pm$0.7    & 23$\pm$0.7            & 7.9$\pm$0.09           & 1.0$\pm$0.01  & 1.0$\pm$0.01  & 4.18$\times 10^{-3}$                                                     & 515                 & 0.0515              & II       \\
PLCK$\_$G213.0+65.9                                        & 16$\pm$17     & 16$\pm$17             & 16$\pm$17              & 11$\pm$0.6    & 3.9$\pm$0.08  & 3.88$\times 10^{-10}$                                                    & 0                   & $< 10^{-4}$         & III (RC) \\
PLCK$\_$G223.9+41.2                                        & 34$\pm$2.5    & 18$\pm$0.2            & 8.5$\pm$0.07           & 4.0$\pm$0.04  & 3.0$\pm$0.03  & 9.97$\times 10^{-4}$                                                     & 365                 & 0.0365              & II       \\
PLCK$\_$G328.9+71.4                                        & 37$\pm$1.9    & 37$\pm$1.9            & 37$\pm$1.9             & 14$\pm$0.2    & 6.9$\pm$0.08  & 8.40$\times 10^{-34}$                                                    & 0                   & $< 10^{-4}$         & III (RC) \\
PLCK$\_$G49.6--42.9                                        & 21$\pm$32     & 7.8$\pm$0.2           & 3.7$\pm$0.03           & 1.1$\pm$0.01  & N/A           & 0.32                                                                     & 1937                & 0.1937              & I        \\
PLCK$\_$G84.0--71.5                                        & 6.6$\pm$0.1   & 6.6$\pm$0.1           & 6.6$\pm$0.1            & 6.6$\pm$0.1   & 3.5$\pm$0.04  & 0.02                                                                     & 700                 & 0.07                & II (C)   \\
PLCK$\_$HZ$\_$G038.0--51.5                                 & 40$\pm$12     & 40$\pm$12             & 12$\pm$0.2             & 6.3$\pm$0.07  & 2.1$\pm$0.02  & 1.17$\times 10^{-6}$                                                     & 87                  & 0.0087              & III      \\
PLCK$\_$HZ$\_$G067.2--63.8                                 & 22$\pm$0.4    & 22$\pm$0.4            & 9.7$\pm$0.09           & 7.3$\pm$0.06  & 4.1$\pm$0.03  & 2.13$\time 10^{-4}$                                                      & 242                 & 0.0242              & III      \\
PLCK$\_$HZ$\_$G103.1--73.6                                 & 15$\pm$0.4    & 15$\pm$0.4            & 7.4$\pm$0.08           & N/A           & N/A           & 4.18$\times 10^{-3}$                                                     & 515                 & 0.0515              & II (R)   \\
PLCK$\_$HZ$\_$G106.8--83.3                                 & 22$\pm$0.4    & 22$\pm$0.4            & 11$\pm$0.1             & 8.5$\pm$0.07  & 2.0$\pm$0.02  & 7.23$\times 10^{-6}$                                                     & 137                 & 0.0137              & III      \\
PLCK$\_$HZ$\_$G119.4--76.6                                 & 24$\pm$0.4    & 24$\pm$0.4            & 7.9$\pm$0.07           & 5.2$\pm$0.05  & 2.0$\pm$0.02  & 4.18$\times 10^{-3}$                                                     & 515                 & 0.0515              & II       \\
PLCK$\_$HZ$\_$G132.6--81.1                                 & 7.5$\pm$0.3   & 7.5$\pm$0.3           & 4.4$\pm$0.05           & 2.1$\pm$0.02  & 2.1$\pm$0.02  & 0.14                                                                     & 1331                & 0.1331              & I (R)    \\
PLCK$\_$HZ$\_$G171.1--78.7                                 & 20$\pm$0.5    & 20$\pm$0.5            & 20$\pm$0.5             & 4.3$\pm$0.04  & 4.3$\pm$0.04  & 4.90$\times 10^{-14}$                                                    & 0                   & $< 10^{-4}$         & III (RC) \\
PLCK$\_$HZ$\_$G173.9+57.0                                  & 7.7$\pm$1.7   & 7.7$\pm$1.7           & N/A                    & N/A           & N/A           & \begin{tabular}[c]{@{}c@{}}0.44\\ (at 6 mJy)\end{tabular}                & 1503                & 0.1503              & I        \\
PLCK$\_$HZ$\_$G176.6+59.0                                  & 12$\pm$0.4    & 12$\pm$0.4            & 12$\pm$0.4             & 12$\pm$0.4    & 5.1$\pm$0.05  & 1.17$\times 10^{-6}$                                                     & 87                  & 0.0087              & III (C)  \\
PLCK$\_$HZ$\_$G214.1+48.3                                  & 14$\pm$8.7    & 14$\pm$8.7            & 14$\pm$8.7             & 11$\pm$0.2    & 11$\pm$0.2    & 2.44$\times 10^{-8}$                                                     & 0                   & $< 10^{-4}$         & III (RC) \\ \hline
\begin{tabular}[c]{@{}l@{}}S2CLS\\ (expected)\end{tabular} & 22.6$\pm$0.34 & $6.3^{+0.16}_{-0.15}$ & $1.97^{+0.09}_{-0.08}$ & 0.61$\pm$0.05 & 0.21$\pm$0.03 &                                                                          &                     &                     &       \\  \hline

\end{tabular}
\end{table*}


We classify these 46 candidate protoclusters into three categories based on their observed source densities over each map area of approximately 0.03 deg$^{2}$, following the method in C19.

(I) Those with an observed source density less than 3 times the expected number from Ge17, equivalently $P_{\textrm{ran}} \geq 5 \times 10^{-2}$ at 8 mJy or $P_{\textrm{ran}} \geq 1.04 \times 10^{-4}$ at 6 mJy, are regarded as not having an overdensity of SCUBA-2 sources.

(II) Those with observed source densities between 3 and 5 times the expected number from Ge17, equivalently $2.13 \times 10^{-4} < P_{\textrm{ran}} < 5 \times 10^{-2}$ at 8 mJy or $1.67 \times 10^{-12} < P_{\textrm{ran}} < 1.04 \times 10^{-4}$ at 6 mJy, are regarded as having an intermediate overdensity of SCUBA-2 sources.

(III) Those with observed source densities greater than or equal to 5 times the expected number from Ge17, equivalently $P_{\textrm{ran}} \leq 2.13 \times 10^{-4}$ at 8 mJy or $P_{\textrm{ran}} \leq 1.67 \times 10^{-12}$ at 6 mJy, are regarded as overdense in SCUBA-2 sources.

We find that there are 25 candidate protoclusters in M17 that can be regarded as being overdense in SCUBA-2 sources (category III, see column 7 in Table \ref{Table_todd}), and are thus the most likely to be bona fide protoclusters. There are 11 candidate protoclusters that contain a mild overdensity of SCUBA-2 sources (category II); these are still likely to be protoclusters, rich in 850-$\mu$m sources but the SCUBA-2 observations may not be sensitive enough to confirm the overdensities. There are also 10 candidate protoclusters in M17 that do not appear to have an overdensity of SCUBA-2 sources (category I); nonetheless, they still have overdensities of \textit{Herschel}-SPIRE sources \citep{2015A&A...582A..30P}. For candidates in all these three categories, photometric data at other wavelengths and/or spectroscopic data will be needed to confirm their protocluster status.

The two candidate protoclusters in M17 having the largest overdensities are PLCK$\_$G328.9+71.4 and PLCK$\_$G160.7+41.0. There are 37 sources above 8 mJy in PLCK$\_$G328.9+71.4 (after sensitivity is corrected over the map) compared to the expected 1.97 sources according to Ge17. This gives an essentially vanishing probability. There are 29 sources above 8mJy in PLCK$\_$G160.7+41.0 (after sensitivity is corrected over the map), which also has negligible probability. Given that 36 out of 46 candidate protoclusters in M17 (78$\% \pm 17 \%$, Poissonian error) can be regarded as strongly or moderately overdense in SCUBA-2 sources, we suggest that the study of M17 selected candidate protoclusters through overdensities of 850-$\mu$m sources similar to the selection in C19.

It is worth noticing that if considering only the PHz-selected candidate protoclusters (with names starting with ``PLCK$\_$HZ" in Table \ref{Table_todd}), the fraction of candidates with significant overdensities (category III) can reach 60$\% \pm 31 \%$, comparable to that of the combined [PHz+PCCS]-selected candidates (54$\% \pm 14 \%$)

\begin{figure}
    \centering
    \begin{subfigure}[b]{0.5\textwidth}
        \includegraphics[width=\columnwidth]{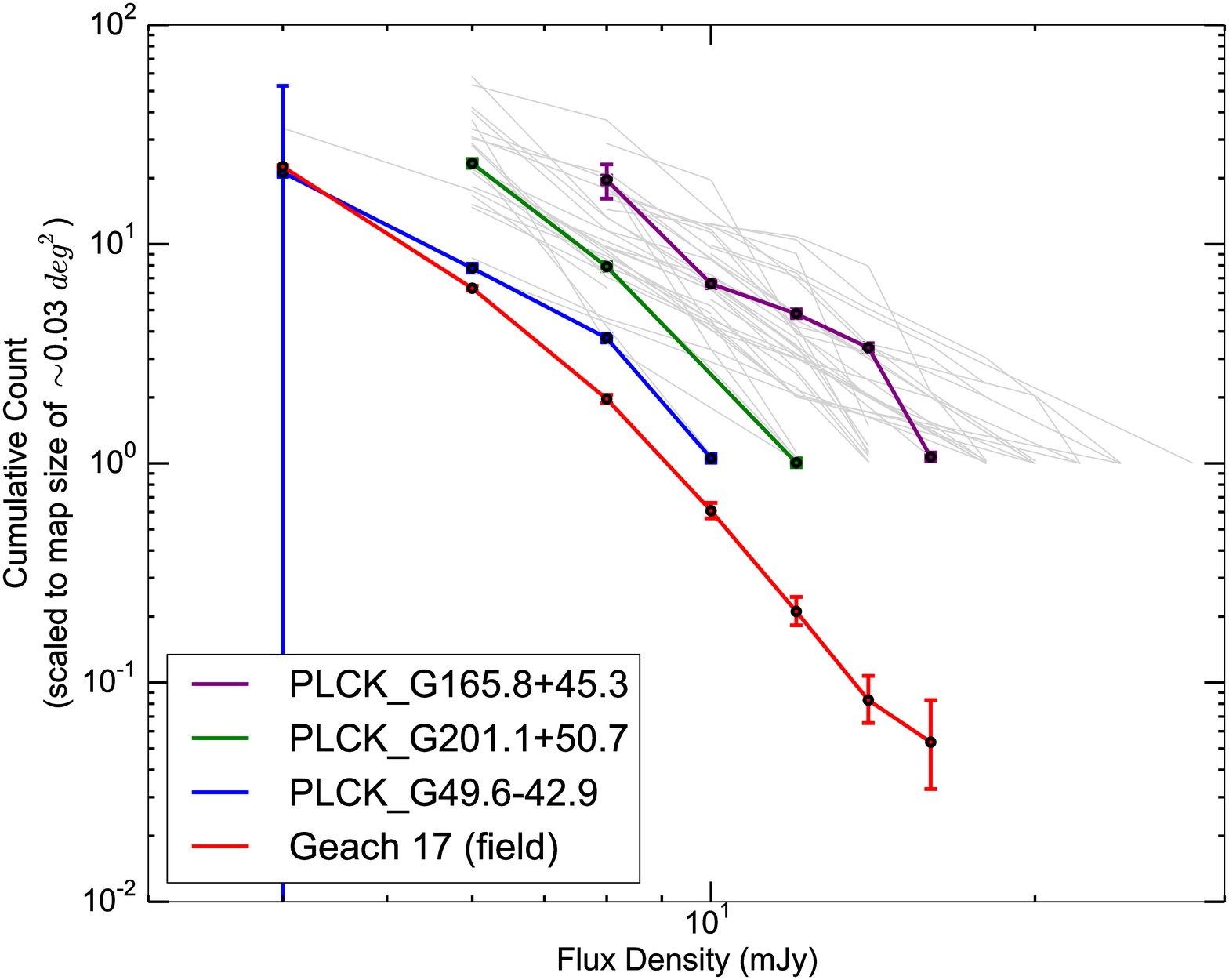}
    \end{subfigure}
    \caption{Cumulative number counts as a function of flux density from S2CLS fields (Ge17, red curve), three example M17 candidate protoclusters from the three categories: (I) no overdensity, blue; (II) intermediate overdensity, green; and (III) significant overdensity, purple. Error bars are Poissonian for Ge17 and completeness for M17 samples. Grey curves show all other M17 candidate protoclusters.}\label{Fig_cum_ncounts}
\end{figure}

Figure \ref{Fig_cum_ncounts} shows the cumulative number counts as a function of 850-$\mu$m flux density for the M17 candidate protoclusters. The red curve shows the cumulative number counts from the S2CLS fields from Ge17, with Poissonian errors. Blue, green, and purple curves show the cumulative number counts of the M17 candidate protoclusters PLCK$\_$G49.6-42.9, PLCK$\_$G165.8+45.3 and PLCK$\_$G201.1+50.7, respectively, with errors propagated from the completeness errors. These candidate protoclusters are representative of categories (I) no overdensity, (II) intermediate overdensity, and (III) significant overdensity, respectively, based on our classification. The grey curves show the cumulative number counts of all other M17 candidate protoclusters. It can be seen that a majority of the candidate protoclusters are overdense compared to the field across a wide range of flux densities.

We estimate the probability that the overdensities are random positive fluctuations due to sample variance \citep{2011ApJ...733...92W, 2012MNRAS.421..284H, 2017MNRAS.464.1380W, 2017MNRAS.470.2253N} instead of real protoclusters following the method in C19. In C19, 10,000 random regions are selected in the map of S2CLS/COSMOS (see Ge17) and the same source extraction algorithm are performed. We count how many have more detected sources than individual candidate protocluster fields ($N_{\textrm{over}}$). The probability of overdensities due to sample variance ($P_{\textrm{over}}$) is simply $N_{\textrm{over}}$/10,000. We also quote $N_{\textrm{over}}$ and $P_{\textrm{over}}$ in Table \ref{Table_todd}.

We find that candidate protoclusters that are in category III (with significant overdensities of 850-$\mu$m sources) have $P_{\textrm{over}} < 3.6 \times 10^{-2}$ and those in category II (with intermediate overdensity) have $P_{\textrm{over}} < 7 \times 10^{-2}$. These results reveal that the overdensities of 850-$\mu$m sources seen in these candidate protoclusters cannot be simply explained by sample variance. The $P_{\textrm{over}}$ values are also consistent with those in C19, who found $P_{\textrm{over}} < 10^{-2}$ for their most overdense candidate protoclusters. In \cite{2016MNRAS.461.1719C}, it was found a $P_{\textrm{over}} = 3.2 \times 10^{-2}$ for H12-00, one of the other candidate protoclusters with submm detections.






\section{Comparison between [PHz+PCCS]- and [ERCSC+PCCS+PCCS2]-selected overdensities}\label{Discussion}

In this study, we compare the 850-$\mu$m number count results of the M17 candidate protoclusters with the 13 candidate protoclusters studied in C19. Those 13 candidate protoclusters were selected originally using the ERCSC, PCCS, and PCCS2 catalogues, and were identified as overdensities of \textit{Herschel} sources in SPIRE bands in \cite{2013A&A...549A..31H}, \cite{2014MNRAS.439.1193C}, and \cite{2018MNRAS.476.3336G}.

Instead of the various colour cuts applied in the M17 catalogue, no colour cut was applied in the ERCSC, PCCS or PCCS2 catalogues on the candidate protocluster selection in the C19 sample. However, when selecting candidate protoclusters with overdensities of \textit{Herschel} sources, \cite{2018MNRAS.476.3336G} applied a 25.4-mJy flux density cut for \textit{Herschel} sources at 350$\mu$m in order to uniformly compare the heterogeneous catalogues from \textit{Herschel}.

We apply the categorisation in Section \ref{todd_ncounts} according to the number counts of 850-$\mu$m sources to classify the 13 candidate protoclusters in C19. Among the candidate protoclusters in C19, five (38$\%$) are in category (I) (no overdensity), four (31$\%$) are in category (II) (intermediate overdensity) and four (31$\%$) are in category (III) (significant overdensity). Among the 46 candidate protoclusters discussed in this study, 10 (22$\%$) are in category (I) (no overdensity); 11 (24$\%$) are in category (II) (intermediate overdensity); 25 (54$\%$) are in category (III) (significant overdensity).

A higher fraction of category (III) candidate protoclusters (with significant overdensities across the same map size) is seen in the sample discussed in this paper. We conclude that the M17 sample, selected from the PHz and PCCS catalogue and with additional colour cuts, has selected a higher fraction of overdensities of 850-$\mu$m sources. Nonetheless, both candidate protoclusters selected in [PHz+PCCS] or [ERCSC+PCCS+PCCS2] can find overdensities of 850-$\mu$m sources at or above the intermediate level (78$\%$ and 62$\%$, respectively).


In addition to the fact that the fraction of candidate protoclusters with significant overdensities is higher in the M17 sample than in the C19 sample, there is evidence that the M17 sample has higher average redshift and infrared luminosity. According to figure 6 of C19 and data from M17, sources from the M17-selected candidate protoclusters have a redshift peak at $3 < z < 4$, whereas the C19 sample has a redshift peak at $2 < z < 3$. In figure 7 of C19, the infrared luminosity of the M17 sample has a peak at $13 < \textrm{log}(L_{\textrm{IR}}(L_{\odot})) < 13.25$ whereas the C19 sample has a peak at $12.75 < \textrm{log}(L_{\textrm{IR}}(L_{\odot})) < 13$. Using source catalogues from M17 and C19, we estimate the mean and standard deviation of redshifts to be $z=3.35 \pm 1.09$ and $z=2.86 \pm 0.96$ for the M17 and C19 samples, respectively. The means and standard deviations of infrared luminosities are $\textrm{log}_{10}(L_{\textrm{IR}} (L_{\odot})) = 13.09 \pm 0.23$ and  $\textrm{log}_{10}(L_{\textrm{IR}} (L_{\odot})) = 12.85 \pm 0.22$ for the M17 and C19 samples, respectively.

We can test if the higher infrared luminosity in the M17 sample is due to them being at higher redshifts, at a fixed flux density at 850$\mu$m. Template SEDs of known DSFGs (local ULIRG Arp220, \cite{2007ApJ...660..167D}, \cite{2011ApJ...743...94R}; average SMGs from the ALMA-LABOCA ECDFS Submm Survey (ALESS), \cite{2015ApJ...806..110D}; the high-\textit{z} source HFLS3 \cite{2013Natur.496..329R}; and the Cosmic Eyelash \cite{2010Natur.464..733S}) are used and their infrared luminosities are estimated from rest-frame 8 to 1,000 $\mu$m at different redshifts, given fixed 850-$\mu$m flux densities. We find that due to the negative K-correction, the infrared luminosity is in general constant at redshifts between $2 < z < 6$. The difference in infrared luminosities between the C19 and M17 samples cannot simply be explained by them being at different redshifts. Hence we also conclude that with additional colour cuts (as discussed in Section \ref{todd_selection}), sources in the [PHz+PCCS]-selected candidate protoclusters (M17 sample) are more luminous and are on average at higher redshifts than sources in candidate protoclusters selected by ERCSC+PCCS+PCCS2 (C19 sample).

In addition to redshifts, M17 also estimated the far-infrared luminosities and star-formation rate densities (SFRDs) of the 850 $\mu$m sources in the candidate protoclusters. They found that the SFRD distribution peaks at a redshift of $z \sim 3$, which is consistent with the peak of the cosmic SFRD in the field \citep{2006ApJ...651..142H, 2012ApJ...754...83B}. We conclude that the M17 sample is robust in selecting 850-$\mu$m source overdensities, most of which are likely to be protoclusters of starbursting galaxies near the peak redshift of the field SFRD at $2<z<3$.

As discussed in Section \ref{todd_selection}, among the 36 category (II) and (III) candidate protoclusters in the M17 sample discussed in this paper, 11 have reliability below 80$\%$ at 3.5$\sigma$ (with ``R" at the ``Category" column in Table \ref{Table_todd}). Adding the fact of the limiting number of sources, there are potential uncertainties when comparing between the M17 and C19 samples as discussed in the last few paragraphs. Future observations are needed to confirm the redshifts and infrared luminosities of these SCUBA-2 sources and their protocluster memberships.


\section{Conclusions}

Forty-six candidate protoclusters were selected in the \textit{Planck} High-\textit{z} catalogue (PHz) and the \textit{Planck} Catalogue of Compact Sources (PCCS), and followed up with \textit{Herschel}-SPIRE and SCUBA-2, as discussed in \cite{2017MNRAS.468.4006M} (M17). We extract sources at 850-$\mu$m using maps from these SCUBA-2 observations with S/N$\geq$3.5, following the method used in \cite{2019MNRAS.490.3840C} (C19).

The cumulative number counts of 850-$\mu$m sources in these cluster candidates are measured from 4 to 12 mJy and the probability ($P$) of the observed number of sources at 8 mJy or 6 mJy is compared with the field values, assuming the sources are randomly distributed. We find that out of 46 candidate protoclusters: 25 have significant overdensities of 850-$\mu$m sources ($P \leq 2.13 \times 10^{-4}$ at 8 mJy or $P \leq 1.67 \times 10^{-12}$ at 6 mJy); 11 have mild overdensities ($2.13 \times 10^{-4} < P < 5 \times 10^{-2}$ at 8 mJy or $1.67 \times 10^{-12} < P < 1.04 \times 10^{-4}$ at 6 mJy); and 10 have no overdensity ($P \geq 5 \times 10^{-2}$ at 8 mJy or $P \geq 1.04 \times 10^{-4}$ at 6 mJy). Approximately 78 percent of the candidate protoclusters have significant or mild overdensities of 850-$\mu$m sources. Hence we conclude that M17, using the PHz and PCCS catalogues, is generally selecting protoclusters with overdensities of 850 $\mu$m sources.

The fraction of candidate protoclusters with overdensities of 850-$\mu$m sources may be underestimated, however, due to the insufficient depth in the M17 survey, for which 21 out of 46 candidate protocluster fields have completeness $<$50$\%$ at 8 mJy.

Comparing this result with the number counts in the C19 sample, which are originally selected from the \textit{Planck} compact source catalogues (ERCSC+PCCS+PCCS2), the [PHz+PCCS]-selected sample has a higher fraction of candidate protoclusters with significant overdensities of 850-$\mu$m sources (54$\%$ versus 31$\%$), has higher photometric redshift and infrared luminosity distributions, due to the additional colour cuts applied. However, the low reliability and small sample size raise some uncertainties when doing these comparisons, which can be improved with future follow-up observations. Nevertheless, both samples show overdensities of 850-$\mu$m sources at or above the intermediate level (78$\%$ for the M17 sample and 62$\%$ for the C19 sample). Hence we conclude that both samples, selected using \textit{Planck} and \textit{Herschel} data, are robust in selecting overdensities of 850-$\mu$m sources, which may be starbursting galaxies in protoclusters near the peak redshift of the cosmic star-formation rate density.

We also want to stress that the confirmation of the [\textit{Planck}+SPIRE+SCUBA-2]-selected targets as genuine overdensities, as discussed in this paper, applies a subset of the bigger PHz list of more than 2,000 sources. There are approximately 10 times of more PHz sources without \textit{Herschel} data, and more without SCUBA-2 data. Hence, follow-up observations using FIR/submm cameras of the rest of the PHz sample in the future may greatly increase the overall cluster/protocluster sample rich in DSFGs and FIR-bright sources, though we cannot totally rule out potential line-of-sight overdensities. Spectroscopic confirmations in the future are needed eventually to rule out this effect.

\section*{Acknowledgements}
The James Clerk Maxwell Telescope is operated by the East Asian Observatory on behalf of the National Astronomical Observatory of Japan, Academia Sinica Institute of Astronomy and Astrophysics, the Korea Astronomy and Space Science Institute, and the Operation, Maintenance and Upgrading Fund for Astronomical Telescopes and Facility Instruments, budgeted from the Ministry of Finance (MOF) of China and administrated by the Chinese Academy of Sciences (CAS), as well as the National Key R$\&$D Program of China (No. 2017YFA0402700). Additional funding support is provided by the Science and Technology Facilities Council of the United Kingdom and participating universities in the United Kingdom and Canada.

The \textit{Herschel} spacecraft was designed, built, tested, and launched under a contract to ESA managed by the \textit{Herschel}/\textit{Planck} Project team by an industrial consortium under the overall responsibility of the prime contractor Thales Alenia Space (Cannes), and including Astrium (Friedrichshafen) responsible for the payload module and for system testing at spacecraft level, Thales Alenia Space (Turin) responsible for the service module, and Astrium (Toulouse) responsible for the telescope, with in excess of a hundred subcontractors.

SPIRE was developed by a consortium of institutes led by Cardiff University (UK) and including: Univ. Lethbridge (Canada); NAOC (China); CEA, and LAM (France); IFSI, Univ. Padua (Italy); IAC (Spain); Stockholm Observatory (Sweden); Imperial College London, RAL, UCL-MSSL, UKATC, and Univ. Sussex (UK); and Caltech, JPL, NHSC, and Univ. Colorado (USA). This development has been supported by national funding agencies: CSA (Canada); NAOC (China); CEA, CNES, CNRS (France); ASI (Italy); MCINN (Spain); SNSB (Sweden); STFC, UKSA (UK); and NASA (USA).


H.D. acknowledges financial support from the Spanish Ministry of Science, Innovation and Universities (MICIU) under the 2014 Ram$\acute{o}$n y Cajal program RYC-2014-15686 and AYA2017-84061-P, the latter one cofinanced by FEDER (European Regional Development Funds).

GDZ gratefully acknowledges financial support from ASI/INAF agreement n.~2014-024-R.1 for the {\it Planck}-LFI Activity of Phase E2 and from the ASI/Physics Department of the university of Roma--Tor Vergata agreement n. 2016-24-H.0


JGN acknowledges financial support from the PGC 2018 project PGC2018-101948-B-I00 (MICINN, FEDER), PAPI-19-EMERG-11 (Universidad de Oviedo) and from the Spanish MINECO for the ``Ramon y Cajal" fellowship (RYC-2013-13256).

E.I.\ acknowledges partial support from FONDECYT through grant N$^\circ$\,1171710.

M.J.M.~acknowledges the support of the National Science Centre, Poland through the SONATA BIS grant 2018/30/E/ST9/00208.

D.R. acknowledges support from the National Science Foundation under grant numbers AST1614213 and AST1910107 and from the Alexander von Humboldt Foundation through a Humboldt Research Fellowship for Experienced Researchers.




\bibliographystyle{mnras}
\bibliography{SCUBA2_overdensities_tcheng_200418} 







\bsp	
\label{lastpage}
\end{document}